\documentclass[12pt]{iopart}

\usepackage{graphicx}
\usepackage{dcolumn}
\usepackage{bm}
\usepackage{bbm}
\usepackage{epsfig}
\usepackage{mathrsfs}
\usepackage{stmaryrd}
\usepackage{color}
\usepackage{dsfont}
\usepackage{iopams}
\usepackage{subeqn}

\begin{document}

\title{Exchange-induced crystallization of soft core bosons}
\author{Fabio Cinti}
\address{Max Planck Institute for the Physics of Complex Systems, 01187 Dresden, Germany}
\address{National Institute for Theoretical Physics (NITheP), Stellenbosch 7600, South Africa}
\address{Institute of Theoretical Physics, University of Stellenbosch, Stellenbosch 7600, South Africa}
\author{Massimo Boninsegni}
\address{Department of Physics, University of Alberta, Edmonton, Alberta, Canada T6G 2E1}
\author{Thomas Pohl}
\address{Max Planck Institute for the Physics of Complex Systems, 01187 Dresden, Germany}
\ead{cinti@sun.ac.za}

\date{\today}

\begin{abstract}
We study the  phase diagram of a two-dimensional assembly of bosons 
interacting via a soft core repulsive pair potential of varying strength, and compare it to that of the equivalent system in which  particles are regarded as distinguishable.  We show that quantum-mechanical exchanges stabilize a ``cluster crystal"  phase in a wider region of parameter space than predicted by 
calculations in which exchanges are neglected.
This physical effect is diametrically opposite to that which takes place in hard core Bose systems such as $^4$He, wherein exchanges strengthen the fluid phase. It is underlain in the cluster crystal phase of soft core bosons by the free energy gain associated to the formation of local Bose-Einstein  condensates. 
\end{abstract}


\maketitle
\section{Introduction.}
The role played by quantum-mechanical exchanges of indistinguishable particles in determining the fluid-solid phase boundary is  a subject of fundamental interest in condensed matter and quantum many-body physics. It has  long been the conventional wisdom  that exchanges should 
have little or no influence over the freezing-melting phase transition. On its face, this assumption would seem reasonable; after all,  in naturally occurring crystals, quantum exchanges are strongly suppressed, both by particle localization at lattice sites, as well as by the strongly repulsive core at short distance of  any  known interatomic or intermolecular potential. Furthermore, melting of most solids occurs at temperatures at which the average rms excursion of particles away from their lattice sites is dominated by thermal effects, quantum-mechanical corrections being generally negligible (see, for instance, Ref. \cite{PhysRevLett.109.025302}).
For this reason, it has  been customary to neglect quantum statistics altogether in theoretical studies of quantum crystals, also near the melting line.

Recent work  \cite{PhysRevLett.109.025302} has challenged this assumption, however, by showing that in Bose systems with hard-core type interactions (such as condensed $^4$He), quantum exchanges have the effect of greatly expanding the region of stability of the fluid phase, with respect to what it would be if exchanges were not present, i.e., if particles were distinguishable (we henceforth refer to distinguishable quantum particles as ``boltzmannons"). To phrase it more quantitatively, the free energy gain associated to the occurrence of long cycles of permutation of identical particles has the effect of moving the freezing line to considerably higher density than one would predict based on calculations only including zero-point motion. For this reason, theoretical studies of the phase diagram of a Bose system neglecting exchanges are likely to incur in significant quantitative error in the determination of the solid-fluid phase boundaries, and to predict unphysical thermocrystallization (i.e., reentrance of the solid phase) at finite temperature \cite{PhysRevLett.109.025302}.
Furthermore, long bosonic exchanges (i.e., comprising a macroscopic fraction of all particles in the system) can  underlie and impart  significant resilience (i.e., long lifetime) to metastable, glassy superfluid phases. Microscopically, this can be phrased in the language of path integrals, in terms of ``frozen"  exchange cycles, in which the paths of many particles become entangled. Because a macroscopic number of single-particle (or, rare multi-particle) tunnelling events are required, in order to disentangle all particles, the system may remain ``stuck" in a metastable disordered, glassy superfluid phase \cite{Boninsegni2006glass}.

In this work, we show that the findings of Ref. \cite{PhysRevLett.109.025302} crucially rely on the presence 
of a ``hard" repulsive core at short distances in the pair-wise interaction $v(r)$. Indeed, if  $v(r)$ features instead 
a ``soft" core (i.e., $v(r\to 0) \sim \hbar^2/2md^2$, where $m$ is the particle mass and $d$ the 
mean inter-particle distance),  the effect of Bose statistics is in fact the {\it opposite}. Specifically, a high density ``droplet" (or, ``cluster") crystal phase, featuring a multiply occupied unit cell \cite{PhysRevLett.105.135301,saccani2011}, is strengthened over the fluid one, again with respect to the physics of a system of boltzmannons (or classical systems featuring the same kind of interaction, e.g., macromolecules \cite{PhysRevLett.109.228301,likos:224502,Likos2001267} \footnote{In these cases, the occurrence of cluster crystal phases can be understood in terms of potential energy alone.}). Phrased differently, since the energy cost associated to   particles laying at 
close distance is relatively small, a  phase in which each unit cell acts in a sense as a mesoscopic Bose condensate, has a lower free energy than the uniform fluid phase\footnote{Quantum-mechanical exchanges may be restricted to particles in the same cell, in which case the crystal is insulating, or particles may hop to adjacent cells and a supersolid phase may ensue, but this aspect is actually not
crucial to the physics of interest here.}.
We arrived at this conclusion by studying a two-dimensional system of soft core bosons by means of quantum Monte Carlo simulations. Although the results presented here are for a specific kind of soft core interaction \cite{PhysRevLett.104.195302,PhysRevLett.106.170401,PhysRevA.87.052314}, experimentally realizable in an assembly of cold atoms \cite{Bloch2008}, the physics described here is independent on  the detailed form of potential utilized in our study, but only on the presence of a soft core at short inter-particle separation \cite{saccani2011}.
\\ \indent
The remainder of this paper is organized as follows: in the next section we describe the mathematical model utilized here. We then
briefly outline the computational methodology, which is fairly standard and extensively documented in the literature, and illustrate our results in the following two sections. We  outline our conclusions in the last section.

\section{Model}

We consider here an ensemble of $N$ spin-zero Bose particles of mass $m$, whose motion is confined to two physical dimensions (a choice made for convenience only, the main physical conclusions being independent of the dimensionality). The system  is enclosed in a square cell of area $A$, with periodic boundary conditions, and is described by the following  many-body Hamiltonian:
\begin{equation}\label{ham2}
\hat H = -\frac{\hbar^2}{2m} \sum_i\nabla_i^2 + \sum_{i<j} v(|{\bf r}_i-{\bf r}_j|)
\end{equation}
The specific form of the potential utilized in this study is
\begin{equation}\label{pot}
v(r) = \frac{v_0}{r_c^6+r^6} 
\end{equation}
with $v_0 > 0$.
Such a potential describes the interaction between two Rydberg atoms \cite{Gallagher1994} in the so-called Rydberg blockaded regime \cite{PhysRevLett.87.037901,Henkel10}. 
The above choice of interaction is motivated by the fact that a quasi-2D Bose assembly with such pair-wise potential can be experimentally realized in an assembly of cold Rydberg atoms \cite{Gallagher1994,Schausz:2012fk,PhysRevLett.107.060402}, which
feature strong van der Waals interactions.\cite{PhysRevLett.110.263201}. They are currently utilized in numerous experiments to study long-range interacting effective spin systems \cite{Schausz:2012fk,ryd_spin3,ryd_spin1,ryd_spin4,ryd_spin2}, as well as for applications in quantum optics \cite{PhysRevLett.105.193603,ryd_opt2,ryd_opt3,ryd_opt4} and quantum information science \cite{ryd_qip1,ryd_qip2,ryd_qip3}.

The most important feature of the potential (\ref{pot}) is the soft repulsive core of radius $r_c$, which is the main consequence of the ``Rydberg blockade" mechanism \cite{PhysRevLett.87.037901}, causing a flattening off of the repulsive part at short inter-particle separation\footnote{Obviously, at a sufficiently short distance the repulsion will start increasing rapidly again, as Pauli exclusion principle prevents electronic clouds of different atoms or molecules from spatially overlapping. The assumption made here is that $r_c$ is much greater than the radius of such inner hard core.}.  The rapidly decaying tail, also repulsive, does not play an important role in the context of this work; indeed, as mentioned in the introduction, the same qualitative  behavior shown here can be observed with a broad class of physical potentials displaying a repulsive 
soft core at short distance, the only requirement being the presence of a negative Fourier component 
\cite{PhysRevLett.109.228301,likos:224502,Likos2001267}. Another important feature is that the strength of the repulsive core of the pair-wise interaction can be ``tuned", allowing one to go from the soft to the hard core regime,  in which qualitatively different physics arise.

We take $\epsilon_0=\hbar^2/mr_c^2$ as our unit of energy (and temperature, i.e., we set the Boltzmann constant $k_B$ to one), and $r_c$ that of length. 
Thus, the density of particles $\rho\equiv N/A$ is expressed  in units of $r_c^{-2}$.
The dimensionless parameter  ${V_0}=mv_0/\hbar r_c^4$ measures the relative 
strength of the interaction compared to the characteristic kinetic energy $\epsilon_0$. \\ \indent
The phase diagram of this system is similar to that of other soft core Bose systems \cite{saccani2011}.
If  $V_0 \to\infty$, the physics approaches that of an ensemble of hard disks, 
whose phase diagram features a low-density fluid (gas), turning superfluid at low temperature,  transitioning at 
sufficiently high density into a  crystalline phase with one particle per unit cell (the presence of a weak repulsive 
tail stabilizes such a crystalline phase at $T=0$ even at low density).
Multiple occupation crystals occur at density $\rho \gtrsim 1$; in this regime, in which no supersolid 
phase is observed, 
the physics of the system in the solid phase mimics that of the Bose Hubbard model \cite{Fisher1989}. 
On the other hand, in a range of values of $V_0$ (roughly  $1 \lesssim V_0\lesssim 20$),
at low temperature the system transitions from the fluid phase directly into a crystalline one featuring multiply occupied sites (unit cells) \cite{PhysRevA.87.061602}. This is the physical regime of interest here.

\begin{figure}[t!]
\begin{center}
\resizebox{0.8\columnwidth}{!}{\includegraphics{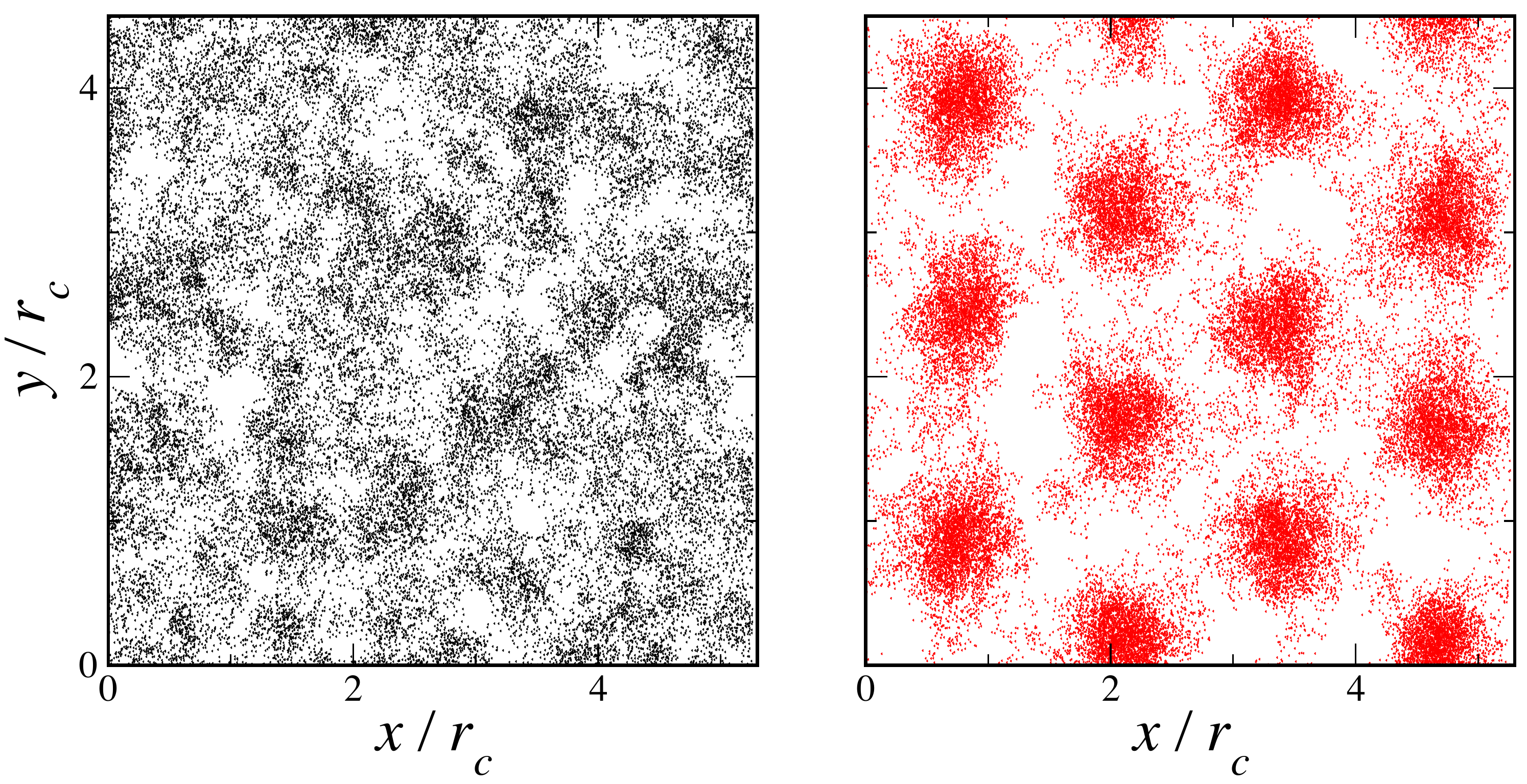}}
\end{center}
\caption{\label{fig1} Configuration snapshots (particle world lines) 
for a system with $N=600$ particles and $\alpha=30$ at a temperature $t=T/\rho=0.3$. Left panel 
refers to a system of distinguishable particles (boltzmannons), the right to one of Bose particles.} 
\end{figure}

\section{Methodology}

We investigated the low-temperature phase diagram of the  system  described by the Hamiltonian (\ref{ham2}) by means of  first principle computer simulations based on the 
 worm algorithm in the continuous-space path integral representation \cite{PhysRevLett.96.070601}. This is a fairly well-established computational methodology, allowing one to obtain essentially exact thermodynamics porperties of Bose systems at finite temperature, using only the microscopic Hamiltonian as input. Because the continuous-space worm algorithm is thoroughly illustrated elsewhere \cite{Boninsegni2006pre}, we shall not review its implementation here. We only mention that details of the calculations are standard, as the use of the potential (\ref{pot}) entails no particular difficulty. We utilized the usual fourth-order approximation for the high temperature density matrix \cite{cuervo}.   All of the results reported here are extrapolated to the limit of zero time step.
\\
The main quantity of interest here is the superfluid density, which we compute by means of the  well-known ``winding number" estimator \cite{pollock87}.
Most of the calculations for which results are shown here were carried out with a number of particles of the order of a few hundred, $N=800$ being the largest system size considered. We carried out parallel simulations, at the same thermodynamic conditions, of a system of boltzmannons described by the same Hamiltonian, in order to assess the effect of Bose statistics on the phase diagram. It is worth reminding that the two systems have the same ground state; this is a straightforward consequence of the fact that the ground state wave function of a many-boson system is nodeless \cite{feynman}. 

\begin{figure}[t!]
\begin{center}
\resizebox{0.8\columnwidth}{!}{\includegraphics{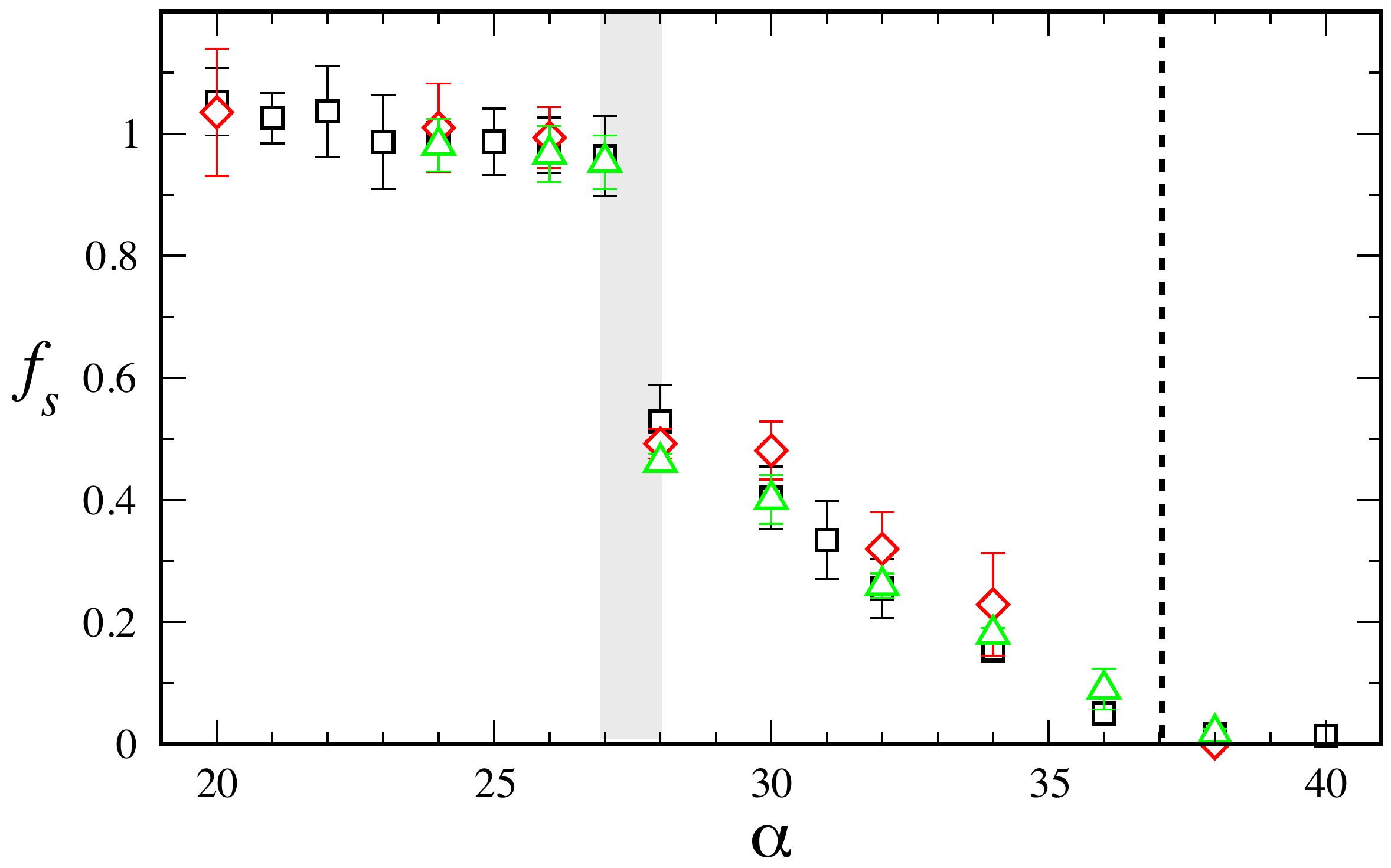}}
\end{center}
\caption{\label{fig2} Superfluid fraction in the limit of $T\to 0$, computed by simulation as a function of 
the renormalized interaction parameter $\alpha=\rho V_0$ (see text), for the three densities
$\rho =$ 4.53 (triangles), 6.78\,(diamonds) and 11.33 (squares). Results shown are for a simulated system comprising $N$=800 particles. When not shown, statistical errors are smaller than the symbols. Shaded area is the coexistence region, as determined from the numerical data. Dashed line indicates the quantum phase transition from a superfluid to a normal cluster crystal.}
\end{figure}

\section{Results}

As mentioned above, the regime of interest in this work is that in which the repulsive core of potential 
(\ref{pot}) is soft, i.e., $V_0 \lesssim 20$. In this range of repulsive interaction,  at a density $\rho \gtrsim 1$ 
the system transitions from a fluid to a droplet crystal phase with a site occupation of the order of a few. 
In particular, considering a density $\rho\approx1$ and with an incommensurate occupation number per site,
it was recently shown as zero-point vacancies cause a superfluid flow of particles through the crystal \cite{cinti}, 
accordingly with the seminal works of  Andreev-Lifshitz-Chester about supersolidity \cite{Andreev,Chester}.
However, in this study we shall consider a number of particles ten or above.

The occurrence of a specific phase, and in particular one that has crystalline long-range order, can be established in a computer simulation by calculating structural quantities like the pair correlation function, which displays marked oscillations in the crystalline phase. Equivalently, its Fourier transform, related to an experimentally measurable quantity 
known as the static structure factor, features a peak in correspondence of the wave vector $k=2\pi/a$, $a$ being the lattice constant (typically $a\sim r_c$).
However, the presence of crystalline order can easily be assessed also by visual inspection of particle world lines, an example 
of which is offered in Fig.~\ref{fig1}, clearly showing the formation of a droplet crystal for a system 
of Bose particles. 
\\ \indent
 A mean-field treatment \cite{PhysRevA.87.061602} shows that the physics of the  
system in the ground state is governed by the single dimensionless parameter $\alpha\equiv V_0 \rho$. 
We have verified by direct numerical simulation that this assertion holds quantitatively for both Bose and 
Boltzmann statistics, at low temperature. 
\begin{figure}[t!]
\begin{center}
\resizebox{0.8\columnwidth}{!}{\includegraphics{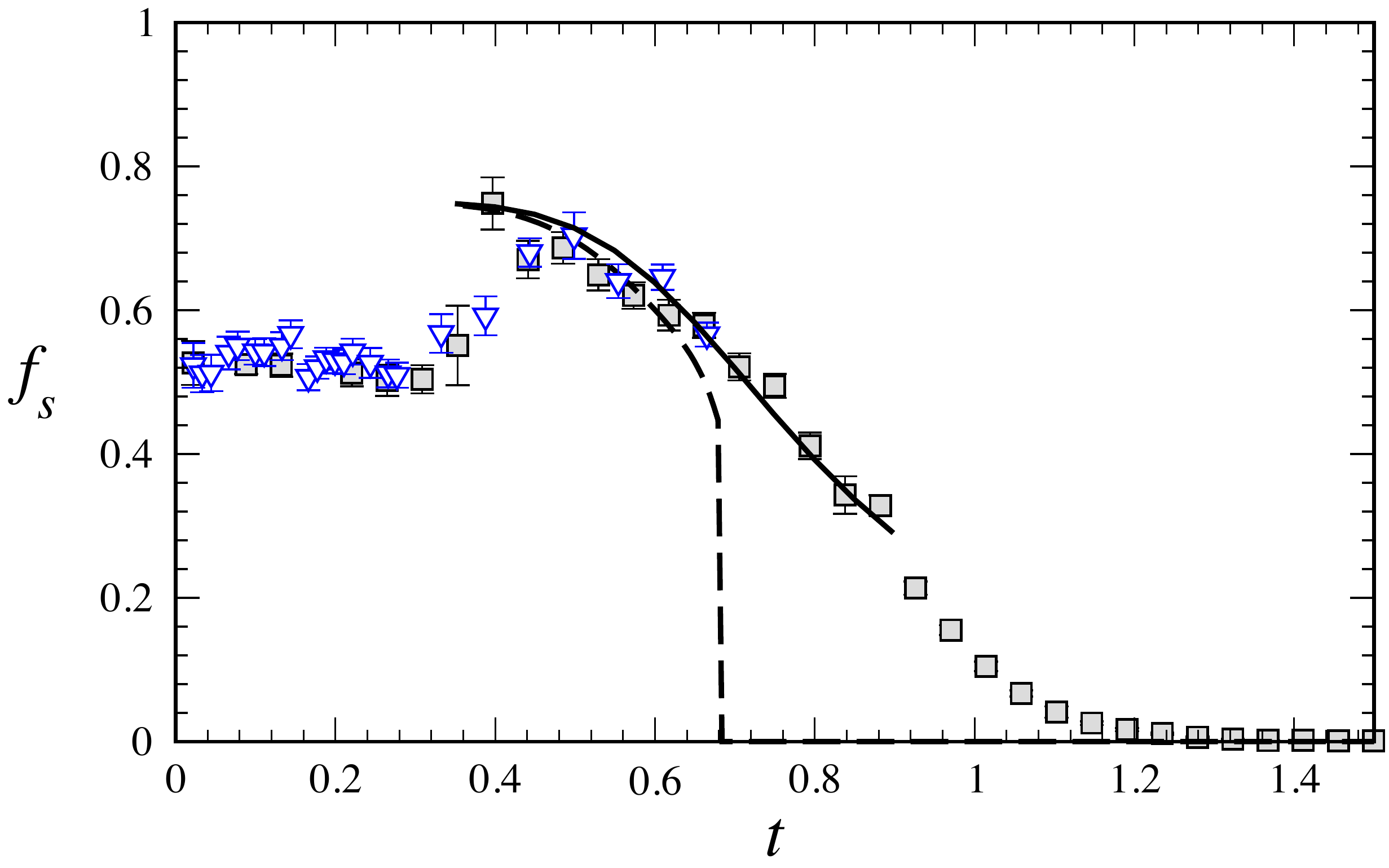}}
\end{center}
\caption{\label{fig3} 
Superfluid fraction versus reduced temperature $t=T/\rho$, for a system with $\alpha=28$, $\rho =$ 11.33 (boxes) and 
$\rho=9.02$ (triangles). The simulated system comprises $N$=800 particles. Line through the data point is a fit based on the 
Berezinskii-Kosterlitz-Thouless theory, whereas the line falling steeply to zero at $t\sim 0.68$ is the prediction for the thermodynamic limit.}
\end{figure}
\begin{figure}[t!]
\resizebox{0.99\columnwidth}{!}{\includegraphics{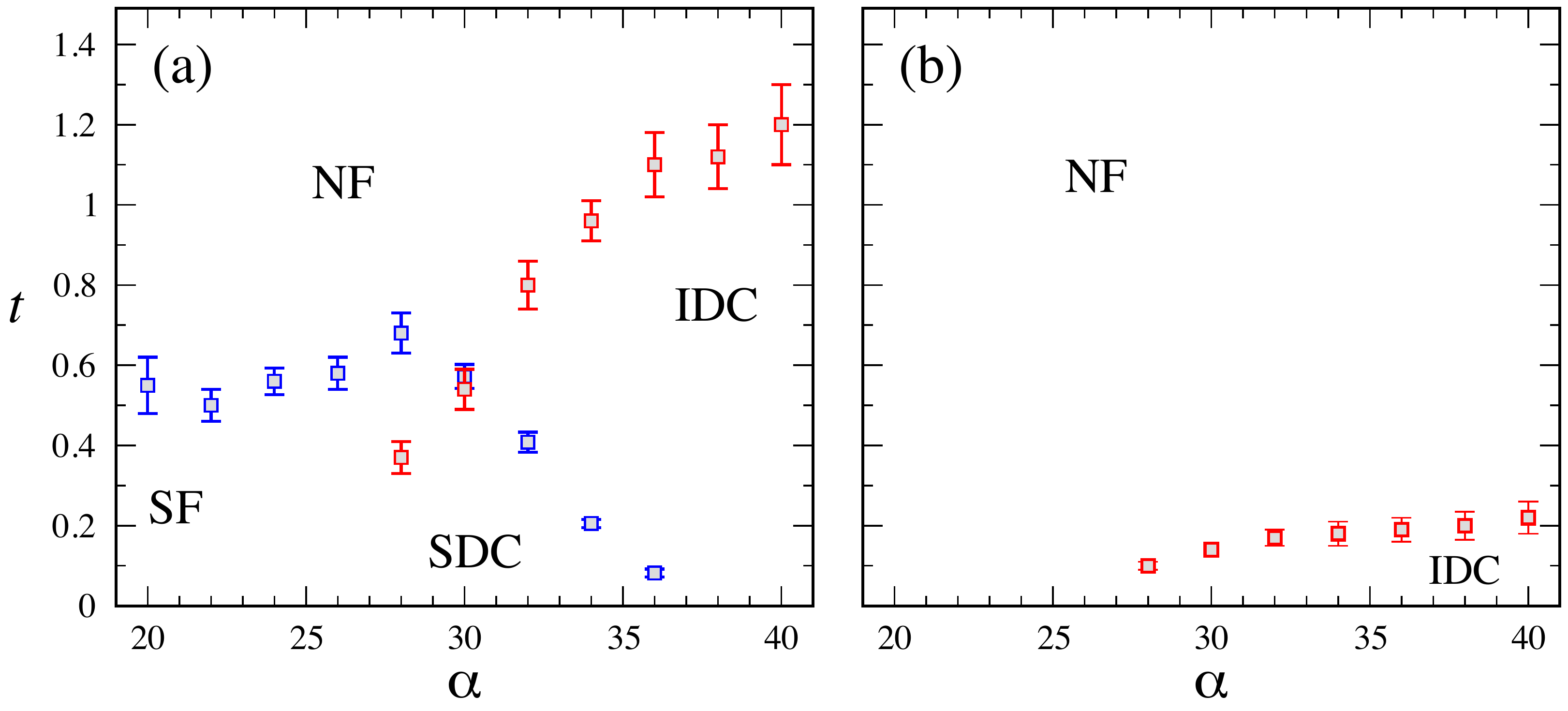}}
\caption{\label{fig4} Phase diagrams in the $\alpha-t$  (see text) plane.
Panel (a): phase diagram for Bose statistics, with 
normal fluid (NF), superfluid (SF), superfluid droplet crystal (SDC), and insulating droplet crystal (IDC) 
Panel (b): phase diagram for a system of distinguishable particles, with NF and IDC phases. 
Symbols represent our numerical determination of the appropriate transition temperatures.}
\end{figure}
To illustrate this point, in Fig.~\ref{fig2} we show results for the superfluid fraction $f_S$ 
computed for the Bose system in the ground state limit (i.e., temperature $T\to 0$) for three different densities, 
namely 4.53, 6.78 and 11.33 (roughly corresponding to 10, 15, and 25 particles per unit cell, respectively).
We show results as a function of the renormalized interaction parameter $\alpha$, 
in a range corresponding to values of the interaction strength $3 \lesssim V_0\lesssim 8$. 
Within the statistical errors of our calculations, the values of the superfluid density all fall on the same curve, 
with three different regimes clearly identifiable. Specifically, at low $\alpha$ the superfluid fraction is $\sim 1$, 
as the system is in the  fluid phase; as $\alpha$ is increased to a value close to 28, $f_S$ abruptly drops to a lower 
(but finite) value, as expected for a superfluid system breaking transational invariance \cite{PhysRevLett.25.1543}.
Finally, as $\alpha$ is increased even further (close to 37), the system transitions into an insulating droplet crystal phase, with negligible particle tunnelling across adjacent unit cells. 
\\ \indent
Fig.~\ref {fig3} shows the superfluid fraction $f_S$ for the Bose system at finite temperature, computed for 
$\alpha=28$, at which value the system displays a supersolid phase at low temperature. We consider here two
values of the density, namely   $\rho=9.02$ and $11.33$; the simulated system comprises $N$=800 particles. We plot the values of $f_S$ as a function of the reduced temperature $t\equiv T/\rho$, and observe collapse of the data. As one can see, $f_S$ starts off at a value slightly less than 0.6 at $t$=0, as the system is in the 
supersolid phase, and jumps up to a higher value at $t\approx 0.4$, in correspondence with the
melting of the crystal into a uniform superfluid. 
The numerical data for $f_S$ in the superfluid regime are fitted in the usual way,   
based on Berezinskii-Kosterlitz-Thouless (BKT) theory \cite{KosterlitzThouless1973,PhysRevB.39.2084}, 
to obtain an estimate of the superfluid transition temperature. 

We construct a schematic phase diagram of both the Bose and the distinguishable particle systems in the 
$(\alpha$, $t$) plane  through a number of vertical ``cuts" at different values of $\alpha$. 
The resulting phase diagrams are shown in Fig.~\ref{fig4}.
The first obvious observation, aside from the fact that no superfluid phase can exist in a system of boltzmannons at any finite temperature,  
and that at exactly $t=0$ the phase diagram is of course the same, is the much greater region of stability of the 
droplet crystal phase in the Bose system. This is the most remarkable outcome of this study, as it runs counter 
to the notion that quantum-mechanical exchanges should in principle favor a uniform phase in Bose systems. 
This is indeed what is observed in numerical studies of hard core Bose systems \cite{PhysRevLett.109.025302}, 
and it is what takes place in the system studied here as well, in the $V_0 \gg 1$ (hard core) limit. 
However, if the repulsive core of the interaction is soft enough that a droplet crystal is present in the phase 
diagram, then quantum-mechanical exchanges of Bose particles actually strengthen the {\it crystalline} phase.  
Thus, Bose statistics stabilizes the crystal phase at a higher temperature 
than in the system of distinguishable particles.

\section{Discussion and Conclusions}
The greater stability of the crystal in the Bose system can be understood in entropic terms. In a  system of distinguishable particles, the crystal melts into a fluid due to the greater entropy of the phase with higher symmetry. On the other hand, in the Bose system at low temperature the entropy of a normal fluid phase in which exchanges are only local in character, is comparable to that of a crystal in which exchanges occur among particles confined to within the same unit cell (droplet). Thus, the thermodynamically preferred phase is that of lower energy, i.e., the crystal. As the strength of the repulsion increases, the number of particles in a droplet decreases and the crystalline phase becomes entropically less competitive with the fluid.

Summarizing, we have shown that in a Bose system characterized by pair-wise interactions with a repulsive core at short distances, quantum-mechanical exchanges can act to stabilize either the fluid or solid phase, depending on the strength of the repulsive interaction. While in the hard core limit exchanges strengthen the fluid phase, the opposite is true in a  system in which the core is soft enough to allow the formation of a cluster (droplet) crystal phase at low temperature.  
We have shown this effect for a two-dimensional system of Rydberg atoms, but the result is quite general, and in particular is independent on the long-range part of the potential.

\section*{Acknowledgements}
We thank Tommaso Macr\`i, Guido Pupillo and Nikolai Prokof'ev for valuable discussions. 
This work was supported in part by the Natural Science and Engineering Research Council of Canada. MB gratefully acknowledges hospitality of the Max-Planck Institute for the Physics of Complex Systems, where the work was carried out.

\vspace{1cm}
\section*{References}
\bibliography{./bose}

\providecommand{\newblock}{}
\begin{thebibliography}{10}
\expandafter\ifx\csname url\endcsname\relax
  \def\url#1{{\tt #1}}\fi
\expandafter\ifx\csname urlprefix\endcsname\relax\def\urlprefix{URL }\fi
\providecommand{\eprint}[2][]{\url{#2}}

\bibitem{PhysRevLett.109.025302}
Boninsegni M, Pollet L, Prokof'ev N and Svistunov B 2012 {\em Phys. Rev.
  Lett.\/} {\bf 109}(2) 025302

\bibitem{Boninsegni2006glass}
Boninsegni M, Prokof'ev N and Svistunov B 2006 {\em Phys. Rev. Lett.\/} {\bf
  96} 105301

\bibitem{PhysRevLett.105.135301}
Cinti F, Jain P, Boninsegni M, Micheli A, Zoller P and Pupillo G 2010 {\em
  Phys. Rev. Lett.\/} {\bf 105}(13) 135301

\bibitem{saccani2011}
Saccani S, Moroni S and Boninsegni M 2011 {\em Phys. Rev. B\/} {\bf 83}(9)
  092506

\bibitem{PhysRevLett.109.228301}
Lenz D~A, Blaak R, Likos C~N and Mladek B~M 2012 {\em Phys. Rev. Lett.\/} {\bf
  109}(22) 228301

\bibitem{likos:224502}
Likos C~N, Mladek B~M, Gottwald D and Kahl G 2007 {\em J. Chem. Phys.\/} {\bf
  126} 224502 (pages~18)

\bibitem{Likos2001267}
Likos C~N 2001 {\em Physics Reports\/} {\bf 348} 267 -- 439 ISSN 0370-1573

\bibitem{PhysRevLett.104.195302}
Henkel N, Nath R and Pohl T 2010 {\em Phys. Rev. Lett.\/} {\bf 104} 195302

\bibitem{PhysRevLett.106.170401}
Maucher F, Henkel N, Saffman M, Kr\'olikowski W, Skupin S and Pohl T 2011 {\em
  Phys. Rev. Lett.\/} {\bf 106}(17) 170401

\bibitem{PhysRevA.87.052314}
Keating T, Goyal K, Jau Y~Y, Biedermann G~W, Landahl A~J and Deutsch I~H 2013
  {\em Phys. Rev. A\/} {\bf 87}(5) 052314

\bibitem{Bloch2008}
Bloch I, Dalibard J and Zwerger W 2008 {\em Rev. Mod. Phys.\/} {\bf 80}(3)
  885--964

\bibitem{Gallagher1994}
Gallagher T~F 1994 {\em Rydberg Atoms\/} (Cambridge University Press,
  Cambridge, England)

\bibitem{PhysRevLett.87.037901}
Lukin M~D, Fleischhauer M, Cote R, Duan L~M, Jaksch D, Cirac J~I and Zoller P
  2001 {\em Phys. Rev. Lett.\/} {\bf 87}(3) 037901

\bibitem{Henkel10}
Henkel N, Nath R and Pohl T 2010 {\em Phys. Rev. Lett.\/} {\bf 104} 195302

\bibitem{Schausz:2012fk}
Schausz P, Cheneau M, Endres M, Fukuhara T, Hild S, Omran A, Pohl T, Gross C,
  Kuhr S and Bloch I 2012 {\em Nature\/} {\bf 491} 87--91

\bibitem{PhysRevLett.107.060402}
Viteau M, Bason M~G, Radogostowicz J, Malossi N, Ciampini D, Morsch O and
  Arimondo E 2011 {\em Phys. Rev. Lett.\/} {\bf 107}(6) 060402

\bibitem{PhysRevLett.110.263201}
B\'eguin L, Vernier A, Chicireanu R, Lahaye T and Browaeys A 2013 {\em Phys.
  Rev. Lett.\/} {\bf 110}(26) 263201

\bibitem{ryd_spin3}
L\"ow R, Weimer H, Krohn U, Heidemann R, Bendkowsky V, Butscher B, B\"uchler
  H~P and Pfau T 2009 {\em Phys. Rev. A\/} {\bf 80}(3) 033422

\bibitem{ryd_spin1}
Schempp H, G\"unter G, Hofmann C~S, Giese C, Saliba S~D, DePaola B~D, Amthor T,
  Weidem\"uller M, Sevin\ifmmode~\mbox{\c{c}}\else \c{c}\fi{}li S and Pohl T
  2010 {\em Phys. Rev. Lett.\/} {\bf 104}(17) 173602

\bibitem{ryd_spin4}
Viteau M, Huillery P, Bason M~G, Malossi N, Ciampini D, Morsch O, Arimondo E,
  Comparat D and Pillet P 2012 {\em Phys. Rev. Lett.\/} {\bf 109}(5) 053002

\bibitem{ryd_spin2}
Hofmann C~S, G\"unter G, Schempp H, Robert-de Saint-Vincent M, G\"arttner M,
  Evers J, Whitlock S and Weidem\"uller M 2013 {\em Phys. Rev. Lett.\/} {\bf
  110}(20) 203601

\bibitem{PhysRevLett.105.193603}
Pritchard J~D, Maxwell D, Gauguet A, Weatherill K~J, Jones M~P~A and Adams C~S
  2010 {\em Phys. Rev. Lett.\/} {\bf 105}(19) 193603

\bibitem{ryd_opt2}
{Dudin} Y~O and {Kuzmich} A 2012 {\em Science\/} {\bf 336} 887--889

\bibitem{ryd_opt3}
{Peyronel} T, {Firstenberg} O, {Liang} Q~Y, {Hofferberth} S, {Gorshkov} A~V,
  {Pohl} T, {Lukin} M~D and {Vuleti{\'c}} V 2012 {\em Nature\/} {\bf 488}
  57--60

\bibitem{ryd_opt4}
Parigi V, Bimbard E, Stanojevic J, Hilliard A~J, Nogrette F, Tualle-Brouri R,
  Ourjoumtsev A and Grangier P 2012 {\em Phys. Rev. Lett.\/} {\bf 109}(23)
  233602

\bibitem{ryd_qip1}
Isenhower L, Urban E, Zhang X~L, Gill A~T, Henage T, Johnson T~A, Walker T~G
  and Saffman M 2010 {\em Phys. Rev. Lett.\/} {\bf 104}(1) 010503

\bibitem{ryd_qip2}
Wilk T, Ga\"etan A, Evellin C, Wolters J, Miroshnychenko Y, Grangier P and
  Browaeys A 2010 {\em Phys. Rev. Lett.\/} {\bf 104}(1) 010502

\bibitem{ryd_qip3}
Saffman M, Walker T~G and M\o{}lmer K 2010 {\em Rev. Mod. Phys.\/} {\bf 82}(3)
  2313--2363

\bibitem{Fisher1989}
Fisher M~P~A, Weichman P~B, Grinstein G and Fisher D~S 1989 {\em Phys. Rev.
  B\/} {\bf 40}(1) 546

\bibitem{PhysRevA.87.061602}
Macr\`i T, Maucher F, Cinti F and Pohl T 2013 {\em Phys. Rev. A\/} {\bf 87}(6)
  061602

\bibitem{PhysRevLett.96.070601}
Boninsegni M, Prokof'ev N and Svistunov B 2006 {\em Phys. Rev. Lett.\/} {\bf
  96}(7) 070601

\bibitem{Boninsegni2006pre}
Boninsegni M, Prokof'ev N~V and Svistunov B~V 2006 {\em Phys. Rev. E\/} {\bf
  74} 036701

\bibitem{cuervo}
Cuervo J~E, Roy P~N and Boninsegni M 2005 {\em J. Chem. Phys\/} {\bf 122}
  114504

\bibitem{pollock87}
Pollock E~R and Ceperley D~M 1987 {\em Phys. Rev. B\/} {\bf 36} 8343

\bibitem{feynman}
See, for instance, R. P. Feynman, {\em Statistical Mechanics: A Set of Lectures}, (Addison-Wesley, New York, 1972).

\bibitem{cinti}
Cinti F,  Macr\`i T, Lechner W, Pupillo G and Pohl T 2014 {\em Nature Comm.} 5:3235 doi: 10.1038/ncomms4235 (2014).

\bibitem{Andreev} 
Andreev A~F and Lifshitz I~M 1969 {\em JETP} {\bf 29} 1107Ð1113

\bibitem{Chester}
Chester  G~V 1970 {\em Phys. Rev. A} {\bf 2} 256Ð258

\bibitem{PhysRevLett.25.1543}
Leggett A~J 1970 {\em Phys. Rev. Lett.\/} {\bf 25}(22) 1543--1546

\bibitem{KosterlitzThouless1973}
Kosterlitz J~M and Thouless D~J 1973 {\em J. Phys. C\/} {\bf 6} 1181

\bibitem{PhysRevB.39.2084}
Ceperley D~M and Pollock E~L 1989 {\em Phys. Rev. B\/} {\bf 39}(4) 2084--2093

\end{thebibliography}
\bibliographystyle{iopart-num}

\end{document}